\documentclass[review,number,sort&compress]{elsarticle}

\usepackage{amssymb}

\usepackage{lineno}
\usepackage{subfig}
\usepackage{float}

\biboptions{sort&compress}

\journal{Nucl. Instru. Meth. A}

\begin{document}

\begin{frontmatter}


\title{Muon Identification Using Deep Neural Networks with the Muon Telescope Detector at STAR}


\author[1,2]{J. D.~Brandenburg$^{*,}$}
\author[1]{Frank Geurts}

\address[1]{Rice University}
\address[2]{Brookhaven National Lab}

\begin{abstract}
The installation of the muon telescope detector opened new possibilities for studying dimuon production at STAR. However, backgrounds from hadron punch-through and weak decays of pions and kaons make the identification of primary muons challenging. In this paper we present a study of shallow and deep neural networks trained as classifiers for the purpose of muon identification using information from the muon telescope detector at STAR. The performance of shallow neural networks is presented as a function of the number of neurons in their hidden layer. A hyperparameter optimization for determining the optimal deep neural network classifier architecture is presented. The optimized deep neural network is compared with shallow neural networks, boosted decision trees, likelihood ratios, and traditional cut-based PID techniques. The superiority of the deep neural network based muon identification technique is demonstrated and compared with traditional PID through the measurement of the $\phi$ meson and the $\psi(2S)$ in p+p collisions at $\sqrt{s}$~=~200 GeV. The deep neural network based PID simultaneously provides higher signal efficiency, signal-to-background ratio, and significance of the $\phi$ peak compared to traditional PID techniques.  Finally, a deep neural network assisted technique for measuring the muon purity in data is presented and discussed.
\end{abstract}

\begin{keyword}

muon identification \sep shallow neural networks \sep deep neural networks \sep multivariate classifiers \sep STAR \sep Muon Telescope Detector
\end{keyword}

\end{frontmatter}

\vspace*{2mm}
\noindent $^*$ Corresponding author. {\it E-mail address:} jbrandenburg@bnl.gov (J. D.~Brandenburg). 
\vspace*{2mm}


\section{Introduction} \label{sec:01introduction}

In 2014 the Solenoidal Tracker at RHIC (STAR) completed its installation of the Muon Telescope Detector (MTD). The MTD has made muon identification over a large momentum range possible for the first time at STAR. However, even with the MTD, identification of pure muons can be challenging due to backgrounds from hadron punch through.  The identification of dimuon pairs is further obscured by secondary muons originating from the weak decays of $\pi\rightarrow \mu + \nu$ and K$\rightarrow \mu + \nu$. We are motivated to explore the possible improvements over traditional techniques in single muon identification and muon pair identification that can be obtained employing modern supervised learning algorithms.

In this paper we explore classification techniques using artificial neural networks (ANN) for improving muon identification using the information provided by the MTD at STAR. In Sect.~\ref{sec:02exp}, a brief description of the relevant STAR subsystems is provided and the variables used for muon identification are defined. In Sect.~\ref{sec:03datasettraining}, the dataset details are provided and the procedure used to generate the training samples is described. In Sect.~\ref{sec:04muonid}, the use of ANN classifiers is explored for muon identification. Both shallow and deep neural networks are compared and the techniques used to determine the optimal deep neural network architecture are discussed and presented. In Sect.~\ref{sec:05results}, the performance of the DNN based muon identification vs. traditional techniques is compared in p+p collisions at $\sqrt{s}$~=~200~GeV. In this section, the use of the trained DNN for data-driven muon purity measurements is also presented. Finally, a summary is presented in Sect.~\ref{sec:06summary}.
\section{STAR Detector} \label{sec:02exp}
The STAR detector is a multi-purpose detector designed with large, uniform acceptance in 0 $< \phi <$ $2\pi$ and $|\eta|<1$. The relevant STAR subsystems used for this study are the Time Projection Chamber (TPC), the Magnet System, the Time-of-Flight (TOF) detector, and the Muon Telescope detector (MTD) \cite{Ackermann2003a,Anderson2003f,Huang2016}. The TPC provides charged particle tracking and particle identification information via ionization energy loss ($dE/dx$) measurement. The TPC sits within a 0.5 T magnetic field, allowing the charge ($q$) and transverse momenta ($p_{T}$) of tracks to be measured from the curvature of their trajectories. The TPC covers $2\pi$ in azimuth and approximately $\pm 1$ unit in pseudo-rapidity ($\eta$) for collisions at the center of the detector. The TPC provides momentum measurement with a momentum resolution of $\sim$1-2\% for muon tracks with 1~GeV/$c$ of momentum at mid rapidity. 

The Time-of-Flight (TOF) detector is installed outside the TPC at a radius of 210 cm and provides precise timing information with a timing resolution larger than $\sim$90~ps in heavy ion collisions\cite{Llope2004}. The TOF detector covers $2\pi$ in azimuth and approximately $\pm 0.9$ unit in pseudo-rapidity. Information from the TOF detector is not used directly for muon identification in this study. Instead the use of timing information from the TOF detector is used in Sec.~\ref{sec:dataset} in the preparation of the labeled training samples. 

\begin{figure}
    \centering
    \includegraphics[width=0.86\textwidth]{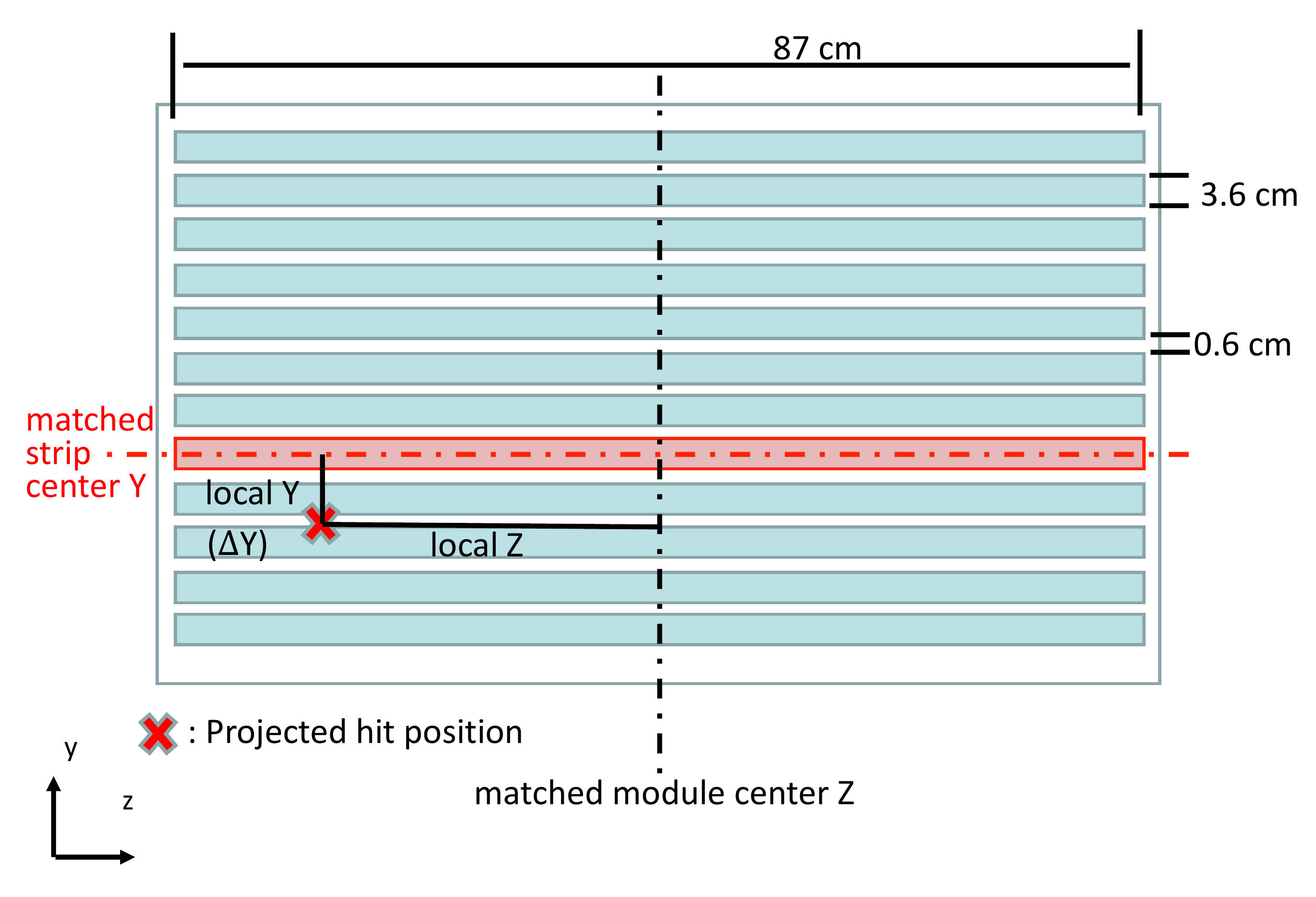}
    \caption{ A schematic of an MTD module. The strips are 87~cm and run along the local $z$ axis. Each module contains 12 strips along the local $y$ axis. Each strip is 3.6~cm wide with a space of 0.6~cm between strips.  }
    \label{fig:mtdschematic}
\end{figure}

The MTD is a multi-gap resistive plate chamber (MRPC) based detector installed outside the magnet's return yoke steel at a radius of 410~cm \cite{Yang2014}. The 60~cm thick magnet steel acts as the hadron absorber, offering up to 5 interaction lengths of material. The MTD is segmented into 30 backlegs in the azimuthal direction each covering about 8 degrees with 2 degrees of gap on either side. On average the MTD covers $\sim$~45\% in the azimuthal direction for $|\eta| <$~0.5. 
The most basic information provided by the MTD for muon identification is whether or not a hit was observed in a given region. In addition to this, the MTD provides precise timing ($\sigma\approx100$~ps) and position ($\sigma\approx2$~cm) measurements which are useful for rejecting hadron punch through. The MTD's double ended strip readout allows the local $Z$ position of hits to be measured via the difference in time between the two ends of the strip. Within each module, the local $Y$ position of a hit is measured by determining which of the 12 strips within a module registered the hit. The $\Delta Z$ and $\Delta Y$ are the residual between the measured local positions and the projected positions in the local $Z$ and $Y$ directions respectively. Figure \ref{fig:mtdschematic} shows a schematic of the local MTD module coordinates and the $\Delta Z$ and $\Delta Y$ calculation. The full list of variables that are used in this study for muon identification are:
\begin{itemize}
    \item $\Delta TOF$ - Difference between the calculated time-of-flight using a muon hypothesis versus the time-of-flight measured by the MTD.
    \item $\Delta Z$ - Difference between the local $Z$ position calculated using a muon hypothesis versus the position measured by the MTD.
    \item $\Delta Y$ - Difference between the local $Y$ position calculated using a muon hypothesis versus the position measured by the MTD from the center of the matched strip. 
    \item cell - the geometric strip index ranging from 0 - 11 with 0 and 11 at the outside edges of each module. The average amount of steel between the interaction point and the MTD module is lowest at the edges.
    \item module - the geometric module index ranging from 0 - 4. 
    \item backleg - the geometric backleg index ranging from 0 - 29. The amount of material between the interaction point and the MTD backlegs varies as a function of backleg since the detector is not fully symmetric in the $\phi$ direction.
    \item n$\sigma_\pi$ - the $dE/dx$ information measured by the TPC. The value normalized by the expected value for the $\pi$ and corrected for detector resolution is used for simplicity. The value of n$\sigma_\pi$ for muons is on average $\sim$+0.5.
    \item DCA - Distance of closest approach of the track to the primary collision vertex.
    \item $p_{T}$ - Transverse momentum of the track. The $\Delta TOF$, $\Delta Y$, and $\Delta Z$ resolutions depend strongly on $p_T$.
    \item $q$ - the track charge measured from its curvature.
\end{itemize}

These variables will be used as the inputs when training neural network classifiers in Sec. \ref{sec:04muonid}.
\section{Dataset and Training Samples} \label{sec:03datasettraining}
\subsection{Dataset and Event Selection}\label{sec:dataset}
The data used for this study was collected by the STAR detector from p+p collisions at $\sqrt{s}$ = 200 GeV during the 2015 RHIC run. The events were selected using the dimuon trigger requiring that at least two MTD signals be measured within a timing window. The primary vertex of the events was required to be within $\pm$ 100 cm of the center of the detector along $z$. In total the dimuon trigger recorded ~300M events corresponding to a total sampled luminosity of 122~pb$^{-1}$ \cite{TAKAHITOQM17}. 

Muon candidate tracks were required to have a $p_T>$~1~GeV/$c$, have a distance of closest (DCA) approach to the collision vertex of $DCA <$~3~cm, be reconstructed from more than 15 clusters in the TPC, have a ratio of reconstructed clusters to possible clusters greater than 0.52 to reject split tracks, and to have at least 10 clusters used for the $dE/dx$ measurement to ensure a reasonable $dE/dx$ resolution. Finally, muon candidate tracks are required to project to active MTD volume and be matched to MTD hits that fired the trigger.  

\begin{figure}
\centering
    \subfloat[]{\includegraphics[width=0.480\textwidth]{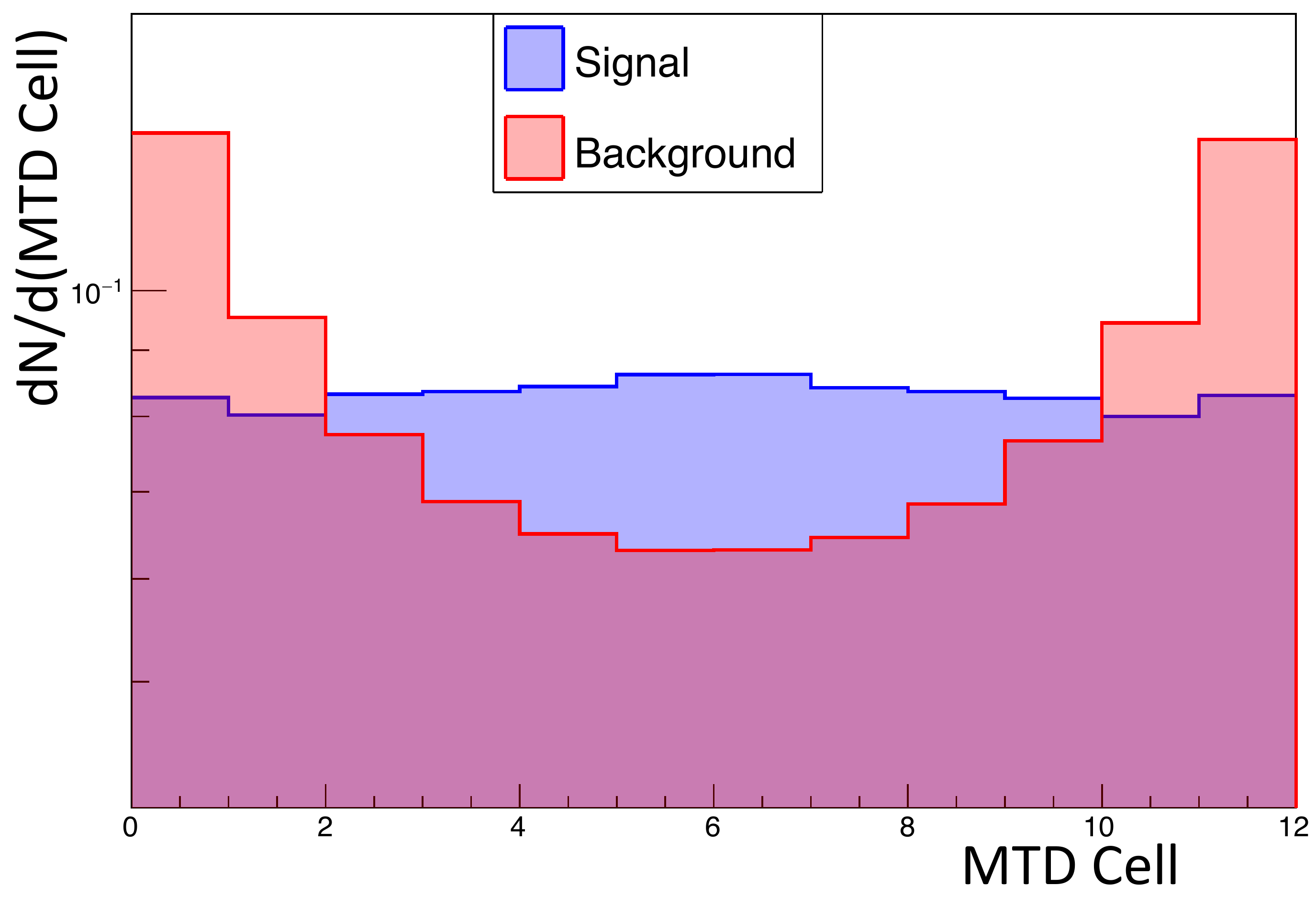}\label{fig:mtdsim-6}}
    \subfloat[]{\includegraphics[width=0.480\textwidth]{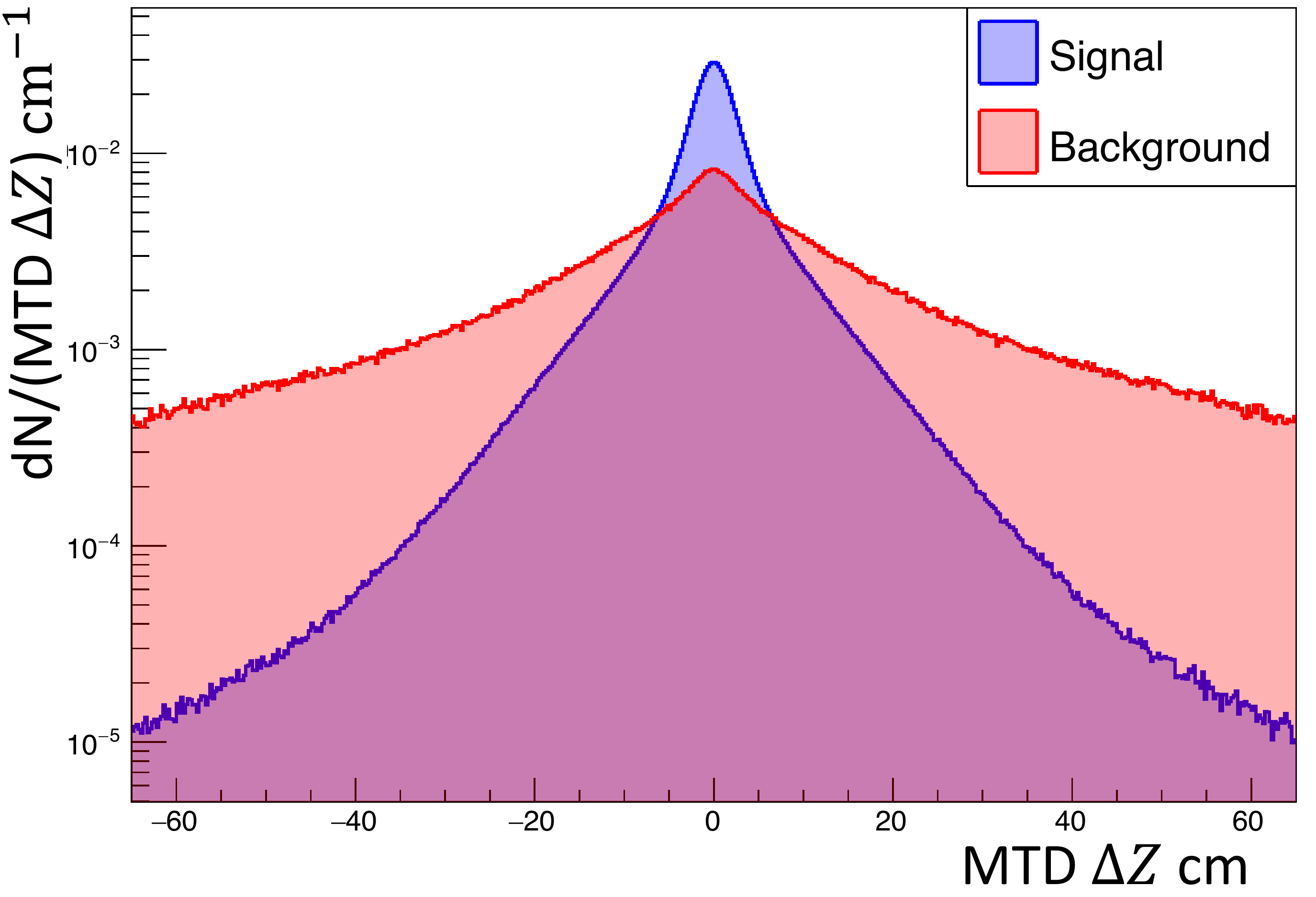}\label{fig:mtdsim-7}}
    \caption{Simulated MTD cell (a) and $\Delta Z$ distributions for signal and background sources. The affect of varying amounts of steel in the $\phi$ direction can be clearly seen in the cell distribution. Hadrons are significantly more likely to punch through to the steel guarding the edge cells (at 0 and 11, respectively) than the central cells. }
    \label{fig:mtdsim2}
\end{figure}

\subsection{ Monte Carlo Simulation} \label{sec:MC}
In Sect.~\ref{sec:04muonid} the training and use of ANNs to perform a two-class classification problem to distinguish signal muons from various types of backgrounds is discussed. This type of ANN based classification is an example of supervised learning and therefore requires labeled datasets for the training phase.  A Monte Carlo (MC) simulation procedure is used to generate the labeled signal and background datasets needed to train the supervised learning algorithms discussed in Sect.~\ref{sec:04muonid}. We define our signal class as primary muon tracks, i.e.\ those originating from the primary interaction vertex. In contrast, the background class includes all other sources of tracks that match to a hit in the MTD and result in a reconstructed track in the tracker. The main sources of background are a result of:

\begin{itemize}
\item punch-through hadrons: e.g., $\pi^{\pm}$, K$^{\pm}$, and $p/\bar{p}$
\item charged-pion weak decays: $\pi \rightarrow \mu + \nu$
\item charged-kaon weak decays: $K\rightarrow \mu + \nu$
\end{itemize}


The procedure used to forward model the signal and backgrounds consists of three main steps: a kinematic event generator, a simulation of the STAR detector, and a full event reconstruction. First, events are generated with $\sim$20 particles/event to mimic the multiplicity of primary tracks in a $p$+$p$ collision at $\sqrt{s}=200$~GeV. Each track in the event is randomly chosen to be a $\mu$, $\pi$, $K$, or $p$. The kinematics of each particle are sampled from flat distributions in $0 < p_{T} < 10.0$~GeV/$c$, $|\eta|<0.8$, and $-\pi < \phi < \pi$. The particle species and kinematics are then fed into a GEANT3 \cite{Brun1987} based simulation of the full STAR geometry. The GEANT3 simulation performs decays of unstable particles, models energy loss of particles traversing media, and interactions with detector materials. Finally, full event reconstruction is performed on the result of the GEANT3 based simulation. This step performs charged particle reconstruction using the simulated hits in the TPC, determines the event's primary interaction vertex, and computes the $dE/dx$ of reconstructed tracks. After tracking is complete the tracks are matched to the simulated MTD hits. The result of this simulation is a set of the PID variables for each of the signal and background processes. An example of the MTD cell and $\Delta Z$ variables are shown for signal and background in Figs.~\ref{fig:mtdsim-6} and \ref{fig:mtdsim-7}, respectively.


\subsection{ Extracting $\Delta$TOF distributions from Data }
A data-driven approach is employed to determine the MTD $\Delta TOF$ distributions separately for the signal and background classes. For this procedure 1D cuts are applied to all PID variables except the $\Delta TOF$. With the cuts listed in Table \ref{table:tof-cuts}, a relatively pure $J/\psi$ sample can be obtained. Figure \ref{fig:dtof_inv_mass} shows the unlike-sign and like-sign distributions near the $J/\psi$ mass after applying the cuts listed in Table \ref{table:tof-cuts}. Daughter tracks from the $J/\psi$ are used to extract the $\Delta TOF$ probability distribution function (PDF) for signal. Specifically the signal PDF is extracted from the $J/\psi$ mass peak (3.0~$<M<$~3.2~GeV/$c^2$) with the background under the peak estimated using the like-sign pairs in the same mass region. The $\Delta TOF$ from the like-sign background is properly scaled and subtracted from the peak region to remove background contributions. The signal $\Delta TOF$ PDF is shown in Fig.~\ref{fig:dtof_inv_mass}. The background $\Delta TOF$ PDF is extracted from tracks passing an inverted set of cuts meant to exclude all signal muons. These cuts are shown in the right hand column of Table~\ref{table:tof-cuts}.

\begin{figure}
	\centering
	\subfloat[]{\includegraphics[width=0.75\textwidth]{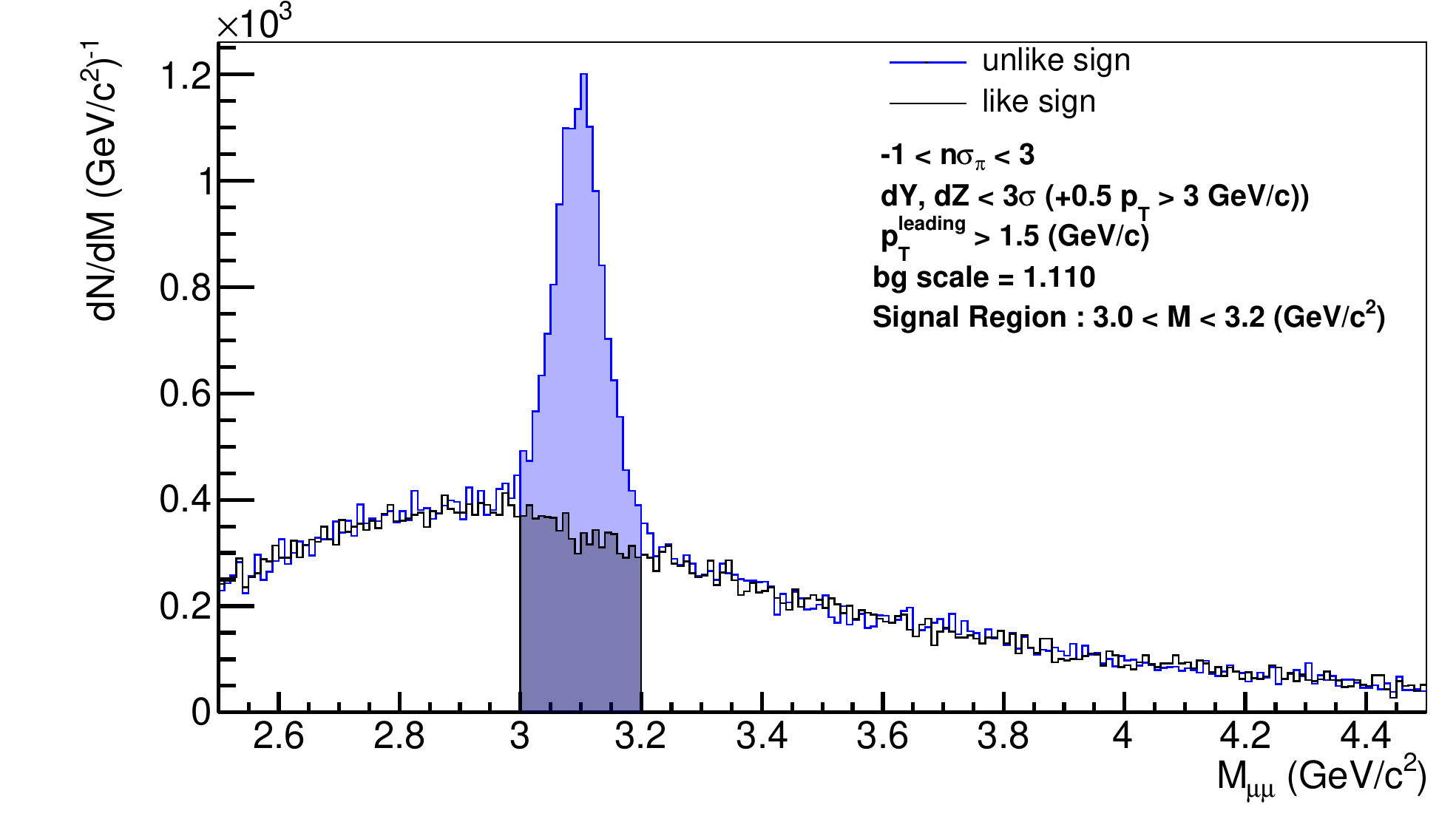}}
	
	\subfloat[]{\includegraphics[width=0.75\textwidth]{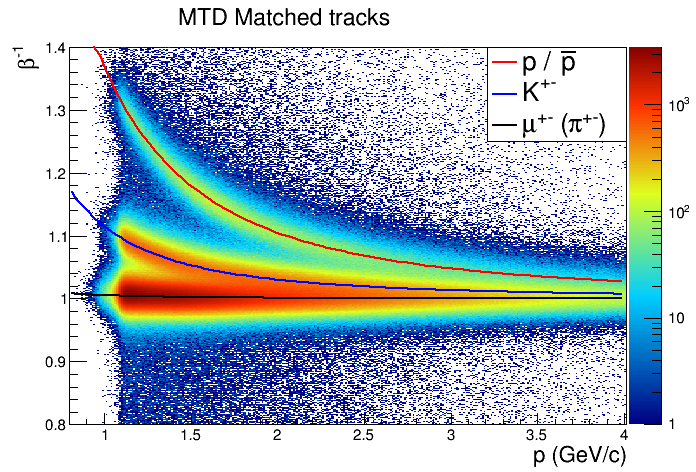}\label{fig:btof-mtd-tracks}}

	\caption{The invariant mass distribution for unlike-sign and like-sign pairs near the $J/\psi$ mass. The $N-1$ cut technique is used to maximize the $J/\psi$ significance by cutting on all MTD PID variables except the $\Delta TOF$ distribution. A $p_T^{\mathrm{leading}} > 1.5$ (GeV/$c$) cut is applied to further improve the purity in the $J/\psi$ mass region. The $\beta^{-1}$ vs.\ momentum distribution for all tracks passing basic QA cuts that are matched to hits in the MTD and the BTOF detectors (b). The $\beta^{-1}$ calculated from the BTOF information shows clear contributions from $\pi$, $K$, and $p/\bar{p}$. }
	\label{fig:dtof_inv_mass}
\end{figure}

\begin{figure}
	\centering
    \includegraphics[width=0.750\textwidth]{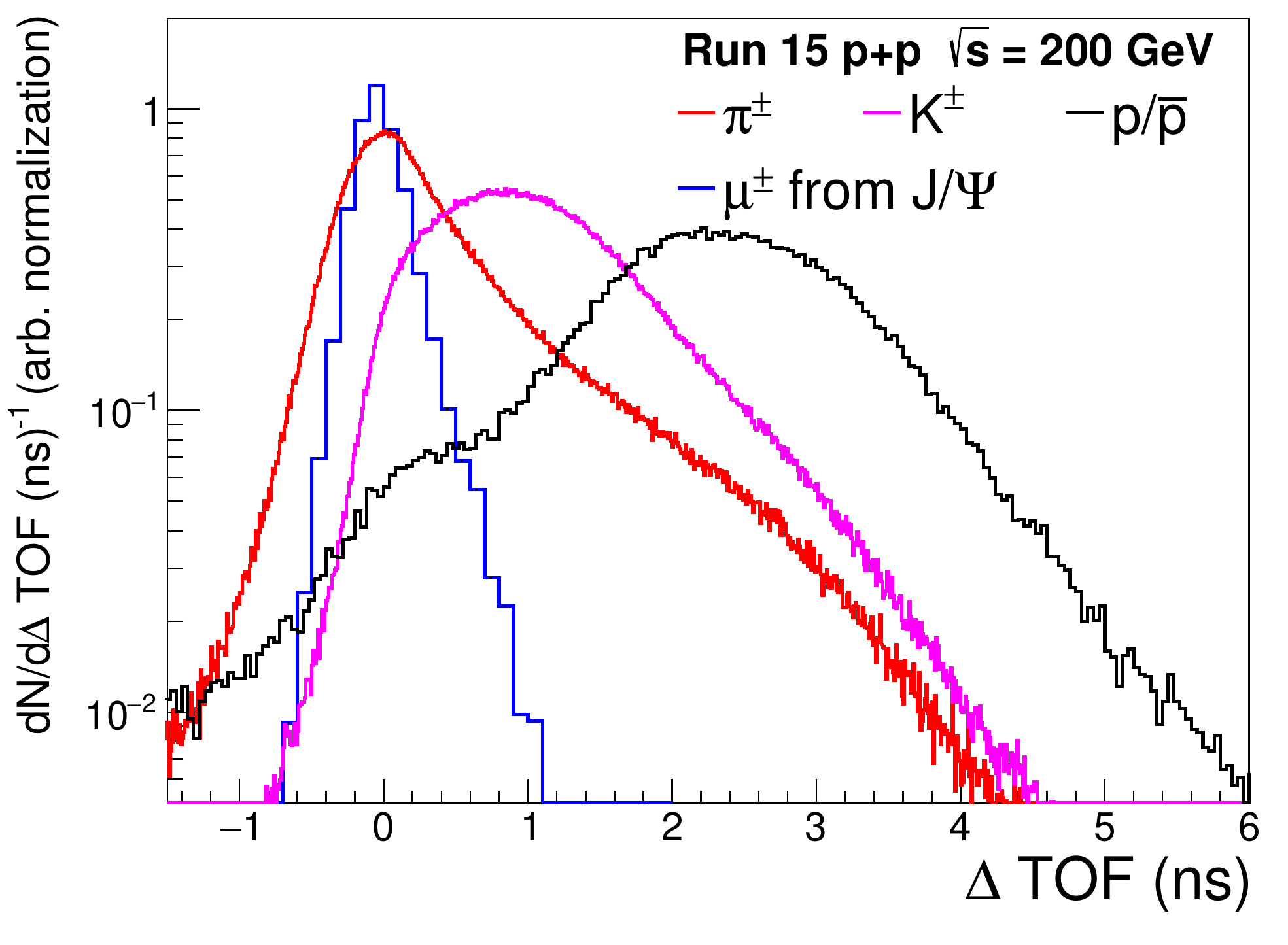}

	\caption{ The $\Delta$TOF distributions for $\mu^{\pm}$ from $J/\psi$, $\pi^{\pm}$, $K^{\pm}$, and $p/\bar{p}$. }
	\label{fig:dtof_inv_mass}
\end{figure}

\begin{table}[]
\centering
\begin{tabular}{l} \hline
$J/\psi$ Selection Cuts                                       \\ \hline \hline
3.0$ < M_{\mu\mu} < $3.2~GeV/$c^2$                       \\ \hline
DCA~$<$~1.0~cm                                                 \\ \hline
-1$ < n\sigma_\pi < $3                                      \\ \hline
$|\Delta$Y$|$ \textless 3$\sigma$ (+0.5, $p_T >$ 3.0 GeV/$c$ ) \\ \hline
$|\Delta$Z$|$ \textless 3$\sigma$ (+0.5, $p_T >$ 3.0 GeV/$c$ ) \\ \hline
$p_{T}^{\mathrm{leading}} > $ 1.5 GeV/$c$                   \\     \hline
\end{tabular}
        \caption{Cuts used for determining the signal and background $\Delta$Time-of-Flight PDFs. }
        \label{table:tof-cuts}
\end{table}

The background $\Delta TOF$ distribution is further separated into the contributions for $\pi$, $K$, and $p$ using timing information from the TOF detector. The sub-sample of tracks which match to both the MTD and TOF are used to extract the MTD $\Delta TOF$ distribution for $\pi$, $K$, and $p$ separately. The $\beta^{-1} = c/v$ distribution measured by the TOF detector is shown in Fig.~\ref{fig:btof-mtd-tracks} for all background tracks matched to both the MTD and TOF. In this figure, there are clear $\beta^{-1}$ bands corresponding to pions, kaons and protons. The MTD $\Delta TOF$ distribution for these three species were extracted by selecting around a given $\beta^{-1}$ band.


    

\subsection{Background MC Closure Test Using Identified $K^0_S \rightarrow \pi^+\pi^-$ and $\phi \rightarrow K^+K^-$ Decays}
\begin{figure}
	\centering
    \subfloat[]{\includegraphics[width=0.48\textwidth]{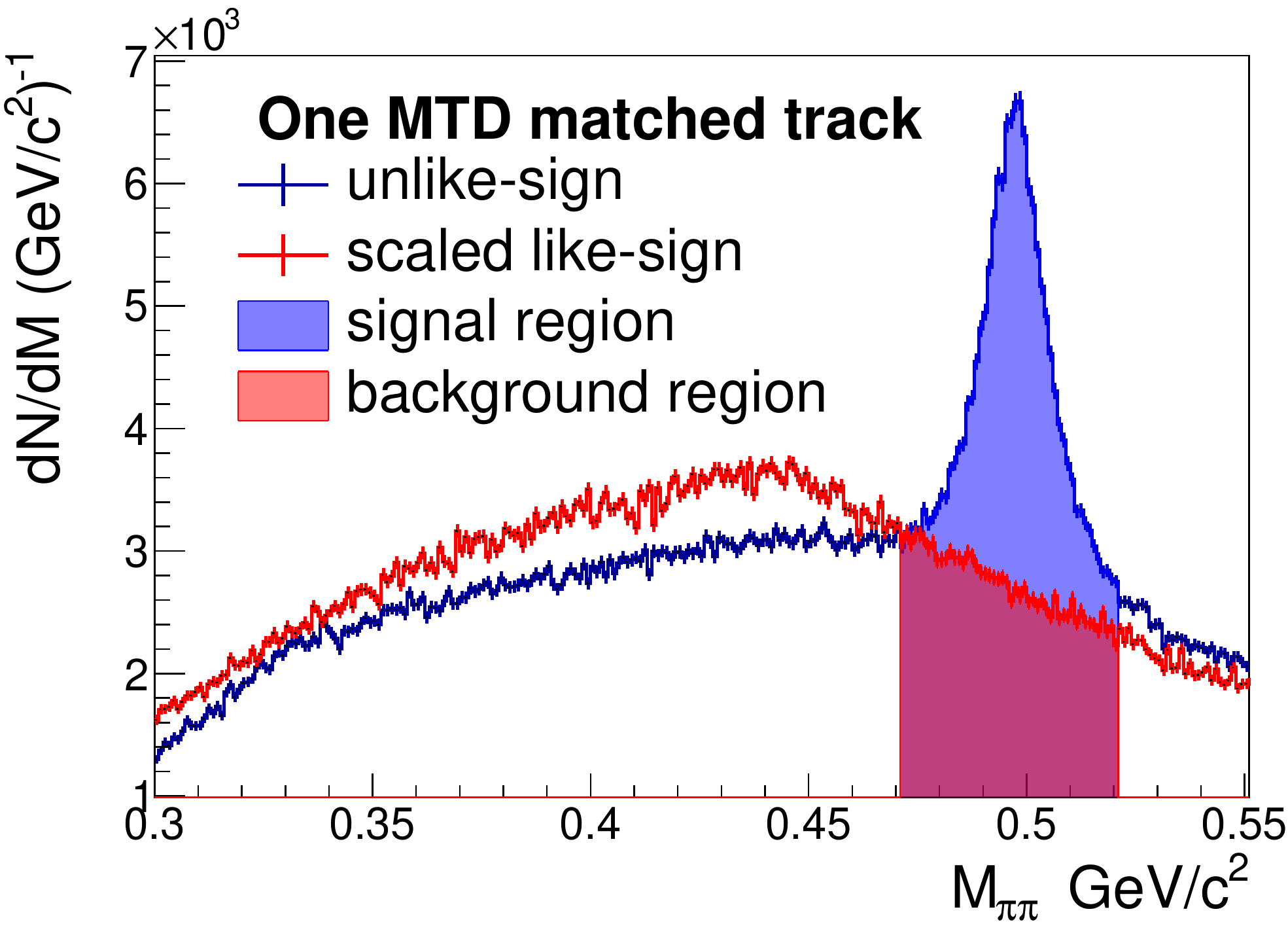}\label{fig:k0spipi}}
    \subfloat[]{\includegraphics[width=0.48\textwidth]{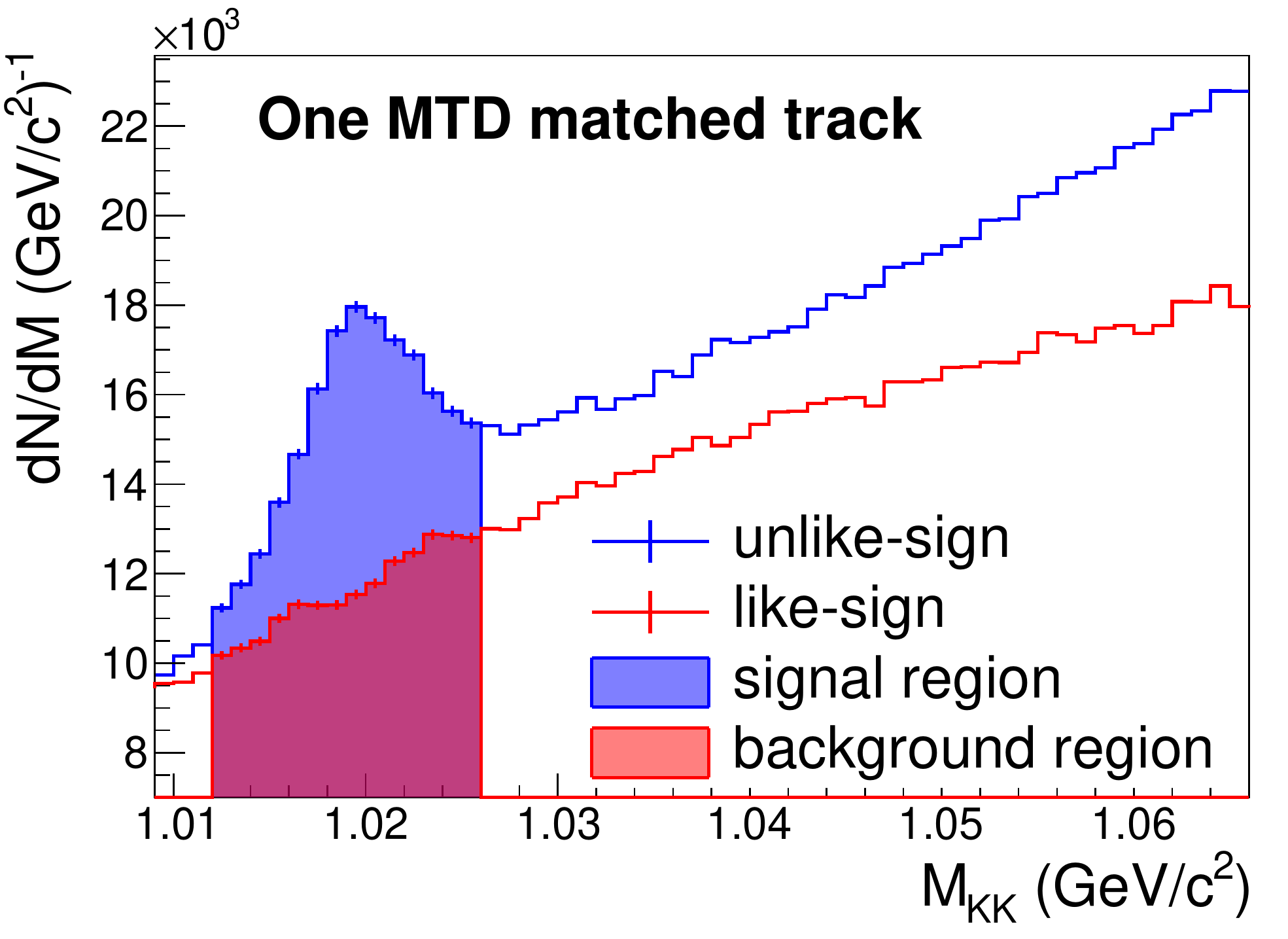}\label{fig:phikk}}
    
    
	 \caption{ The M$_{\pi^+\pi^-}$ distribution near the $K^0_S$ mass shown for the cases in which only one track is matched to an MTD hit (a) and the M$_{K^+K^-}$ distribution near the $\phi$ mass shown for the cases in which only one track is matched to an MTD hit (b).  }
	\label{fig:mc-closure}
\end{figure}

Selecting $K^0_S \rightarrow \pi^+\pi^-$ decays in data provides a $\pi^{\pm}$ enhanced sample that can be used to test the validity of the MC simulation procedure for the $\pi^{\pm}$ background sources. The selection of $K^0_S$ candidates is carried out by applying the topological selection cuts listed in Table~\ref{table:k0s-cuts}. In order to increase the available statistics for comparison only one of the $K^0_S$ daughters is required to have a matching hit in the MTD. Figure~\ref{fig:k0spipi} shows the $\pi^+\pi^-$ invariant mass distribution near the $K^0_S$ mass used to select $\pi^{\pm}$ daughter tracks. The $\pi^{\pm}$ $\Delta Y$, $\Delta Z$, and cell distributions are computed using the unlike-sign distribution minus the scaled like-sign distribution for each variable in the $K^0_S$ mass region (497~$\pm$~25~MeV/$c^2$). 

\begin{figure}
	\centering
    
    \subfloat[]{\includegraphics[width=0.48\textwidth]{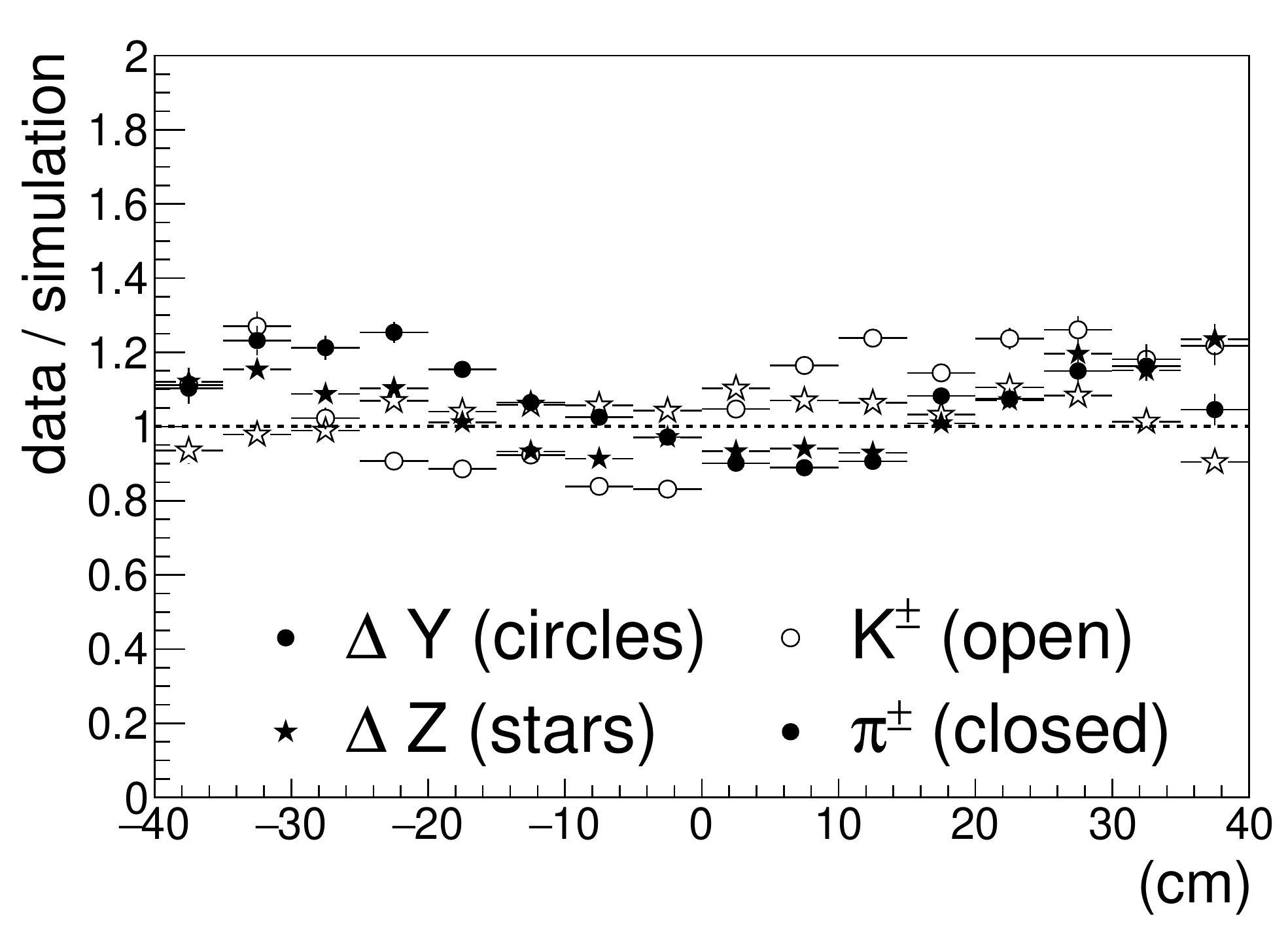}\label{fig:closure-1}}
    \subfloat[]{\includegraphics[width=0.48\textwidth]{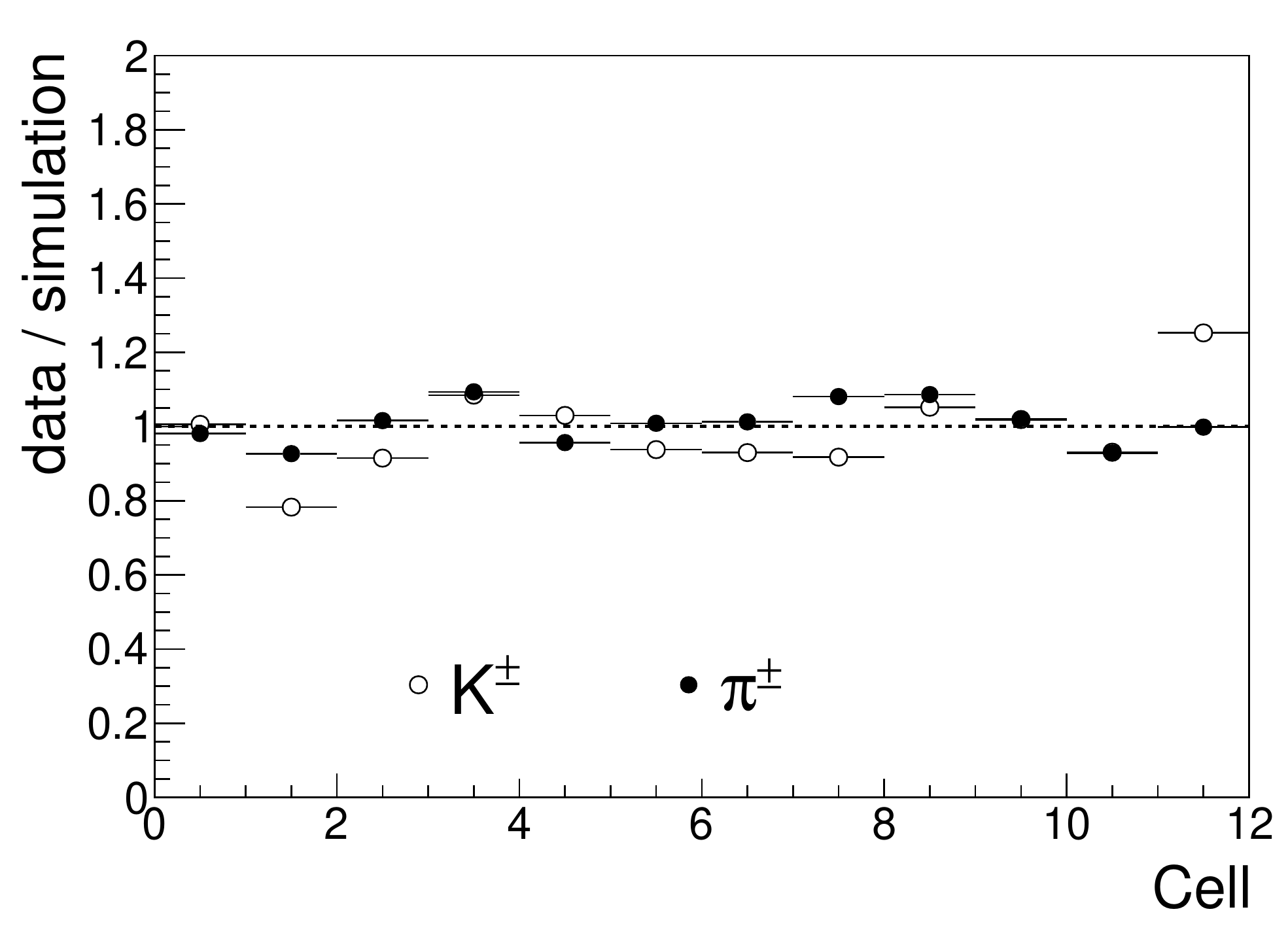}\label{fig:closure-2}}
    
	 \caption{ The $\Delta Y$ and $\Delta Z$ data / simulation ratio for both $\pi^{\pm}$ and $K^{\pm}$ (c). The MTD cell data / simulation ratio for both $\pi^{\pm}$ and $K^{\pm}$ (d).  }
	\label{fig:mc-closure}
\end{figure}

Distributions with an enhanced kaon yield can be selected from the daughters of $\phi \rightarrow K^+K^-$ decays. The $K^+K^-$ invariant mass distribution around $M_{\phi}$ is shown in Fig.~\ref{fig:phikk} for the case in which one track is matched to an MTD hit. The $K^{\pm}$ $\Delta Y$, $\Delta Z$, and cell distributions are computed using the unlike-sign distribution minus the scaled like-sign distribution for each variable in the $\phi$ mass region (1.019~$\pm$~0.007 MeV/$c^2$). The comparison between the $\Delta Y$, $\Delta Z$ and MTD cell distributions from MC and data for $\pi^{\pm}$ and $K^{\pm}$ tracks are shown in Figs.~\ref{fig:closure-1} and \ref{fig:closure-2}. The data / simulation ratios show that the $\Delta Y$, $\Delta Z$ and MTD Cell distributions agree within $\sim\pm$20\%.

\begin{table}[]
\centering
\begin{tabular}{l}
\hline
$K^0_S$ Selection Cuts                            \\ \hline \hline
0.472$ < M_{\pi\pi} < $0.522 GeV/$c^2$            \\ \hline
decay length $>$ 2.7~cm                           \\ \hline
daughter mutual DCA $<$ 1.5~cm                    \\ \hline
pointingAngle $<$ 0.162+0.1123$p_T$ +0.025$p_T^2$ \\ \hline
$|n\sigma_\pi| <$  3                              \\ \hline
\end{tabular}
        \caption{Cuts used to select $K^0_S\rightarrow \pi^+\pi^-$ decays. The daughter pions provide a $\pi$-enhanced sampled that can be compared to the $\pi$ MonteCarlo simulation. }
        \label{table:k0s-cuts}
\end{table}

\section{ Training and Evaluation of Neural Networks } \label{sec:04muonid}

In this section the use of dense Multilayer Perceptrons (MLP), a type of feed-forward ANN, are trained as continuous classifiers for the purpose of muon identification. First, shallow artificial neural networks (SNN) will be discussed. A shallow artificial neural network is defined by the presence of a single hidden layer of neurons between the input and output layers. The universal approximation theorem ~\cite{Cybenko1989,Hornik1989} states that a feed-forward ANN with certain activation functions and at least one hidden layer containing a finite number of neurons can approximate any continuous function on compact subsets of $\mathbb{R}^n$. However, the universal approximation theorem makes no claim about the size of the hidden layer required to approximate a given function. In practice the number of neurons in the hidden ($N_{H}$) layer may need to be intractably large to approximate the desired function with acceptable error. In addition, with increasing number of neurons the risk of over training can increase resulting in a model capable of representing the input data with small error but with very poor generalization performance.

\begin{figure}
	\centering
    \includegraphics[width=0.450\textwidth]{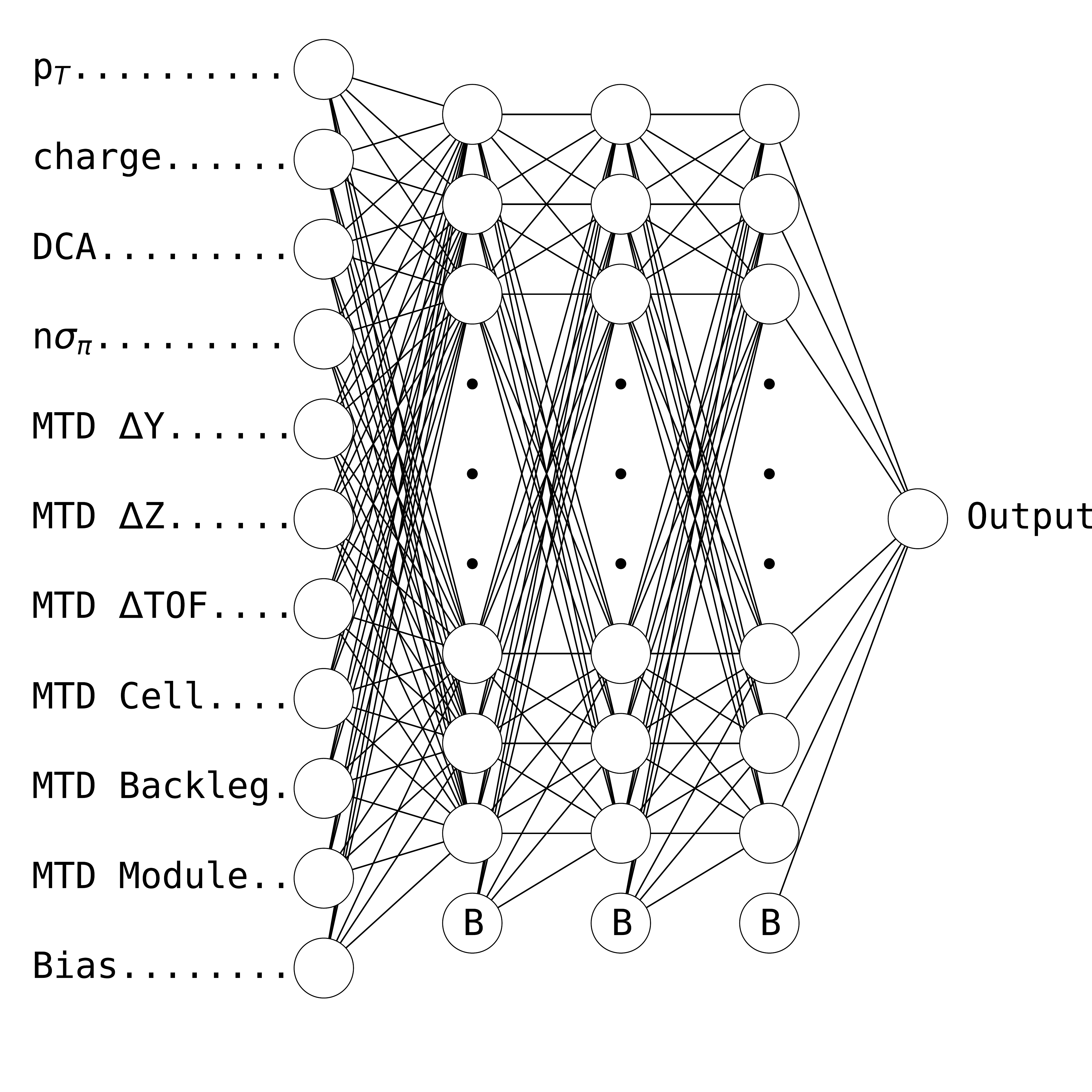}
	\caption{ An example of a dense multilayer perceptron neural network architecture. The shallow neural networks have only a single hidden layer of neurons between the input and output layers. The deep neural networks have two or more. Bias neurons in the hidden layers are marked with a "B". }
	\label{fig:snn}
\end{figure}

\begin{figure}
	\centering
    \includegraphics[width=0.750\textwidth]{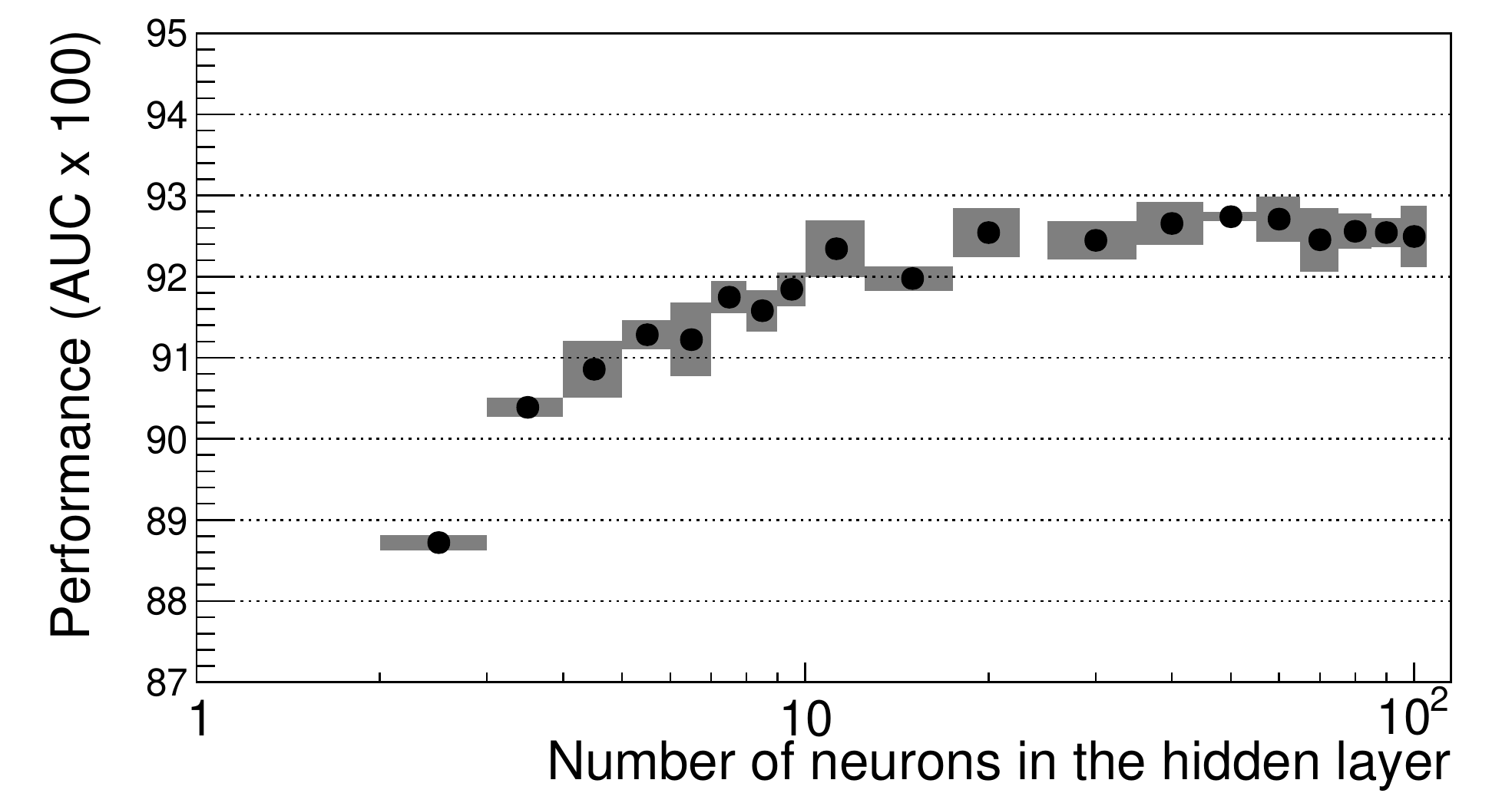}
	\caption{ The signal vs.\ background rejection power as a function of the number of neurons ($NN_{HL}$) in the hidden layer of a shallow neural network. The performance of the SNNs are quantified using the AUC - the area under the background rejection vs.\ signal efficiency curve (See \ref{sec:0501mvacomp} in text). The points are the mean value of 10 models trained with different random samples. The uncertainties show the $\pm 1\sigma$ of the models assuming a Gaussian variance.  }
	\label{fig:snn}
\end{figure}

\subsection{Shallow Neural Networks} \label{sec:snn}
In this section an exploration of the performance of a large set of SNNs as a function of the number of neurons in their hidden layer is presented. The models are trained using the Toolkit for Multivariate Data Analysis with ROOT (TMVA) \cite{Hoecker2007}. Table~\ref{tbl:snnparams} lists the parameters used in the training phase for all models. Each model is trained on a random subset of 100K signal events and 100K background events. A disjoint testing sample is drawn from 250K signal and 250K background events to test the model's response and to evaluate the over training score. The use of a Monte Carlo generator for producing the labeled training samples allows an essentially unlimited number of labeled data sets and allows independent samples for training and testing. When an unlimited labeled data set is not available, bootstrapping techniques can be used to evaluate model performance \cite{Efron1987,Efron1979}. The SNN models include a bias neuron in the each of the input and hidden layers to account for trivial offsets in the mean value of the data. The bias neuron is always "on" - i.e.\ it provides an input of 1 so that weights between it and other neurons are constant factors. The use of a bias node in the input and each hidden layer has become standard practice in neural network architecture design.

Shallow neural networks were trained with 1 to 500~neurons in the hidden layer. For each value of $N_H$, 10~models were trained with different randomized training and testing samples. The performance of each trained SNN is quantified using the area under the curve (AUC) of the background rejection versus signal efficiency distribution (higher is better). The results of the SNN scan are summarized in Fig.~\ref{fig:snn} where the AUC is shown as a function of $N_H$. Each point shows the mean response of 10 models with uncertainties that show the 1$\sigma$ variation between the response of the 10 models assuming a Gaussian variance. The background rejection power of the SNN shows clear improvement as $N_{H}$ is increased until $N_{H}\approx30$. Above $N_{H}\approx30$, adding more and more neurons provides relatively smaller and smaller improvement in the background rejection power. 

\begin{table}[]
\centering
\caption{Parameters used in the training phase for the shallow and deep neural networks.}
\label{tbl:snnparams}
\begin{tabular}{p{4cm} p{2.5cm}}
\hline
Parameter                  & Value            \\ \hline \hline
Neuron Activation Function & tanh             \\
Estimator Type             & Mean Square      \\
Neuron Input Function      & sum              \\
Training Method            & Back-Propagation \\
Learning Rate              & 0.02             \\
Decay Rate                 & 0.01             \\
Learning Mode              & Sequential       \\
Max \# Training Cycles     & 500              \\
Testing Rate               & 100              \\ \hline
\end{tabular}
\end{table}

\subsection{Deep Neural Networks and Hyperparameter Optimization} \label{sec:dnn}
Deep neural networks, in contrast to SNNs which contain only a single hidden layer, contain two or more hidden layers. The additional hidden layers can allow a network to learn complex relationships between input features with far fewer neurons and connections than a shallow network would need. Depending on the application it is also common for DNNs to combine various types of layers, such as convolutional layers, to promote the learning of specific types of relationships. The term, ``Deep Learning'' is often used when a DNN contains several hidden layers with varying types of relationships. In this case, only the simplest type of DNN is explored, specifically dense multilayer perceptrons. Figure~\ref{fig:snn} shows an example of a dense MLP neural network with three hidden layers.

For the case of muon identification, the goal is to determine if DNNs can provide a better classification performance than SNNs with a reasonable number of neurons. Answering this question is not trivial though, since the performance and response of a deep MLP can depend strongly on the number of hidden layers and the number of neurons in each layer. The process of determining the optimal DNN architecture is often referred to as hyperparameter optimization. A grid-search strategy is used in this case to search the optimal DNN architecture on a grid of the number of hidden layers and the number of neurons in each layer. The order of the hidden layers was also encoded so that a network with hidden layers HL$=5, 6, 7$ (i.e.\ 5 neurons in the first hidden layer, 6 in the second, and 7 in the third) would be a distinct grid-point compared to one with HL$=7,6,5$ despite having the same number of hidden layers and number of neurons. For each grid-point a DNN was trained and evaluated based on the following criteria:
\begin{itemize}
    \item Signal vs.\ background rejection power
    \item Prefer simplest NN architecture (fewer number of neurons is better and fewer number of hidden layers is better)
    \item Prefer monotonically increasing S/B as a function of NN response
\end{itemize}
These three criteria are considered to determine the optimal set of DNN hyperparameters. Each DNN was trained using the parameters listed in Table~\ref{tbl:snnparams} with only the architecture related parameters varying. Training DNNs can require significantly more time and larger labeled samples compared to SNNs to reach convergence. The DNNs were trained with 1M signal and 1M background events and took between 10 and 100 times longer to train than the set of SNNs depending on the specific architecture. However, the time-cost required to train DNNs can be greatly reduced by employing modern libraries like TenserFlow that have been heavily optimized for parallelized network training using GPUs \cite{Abadi}.


\begin{figure}
	\centering
    \includegraphics[width=0.85\textwidth]{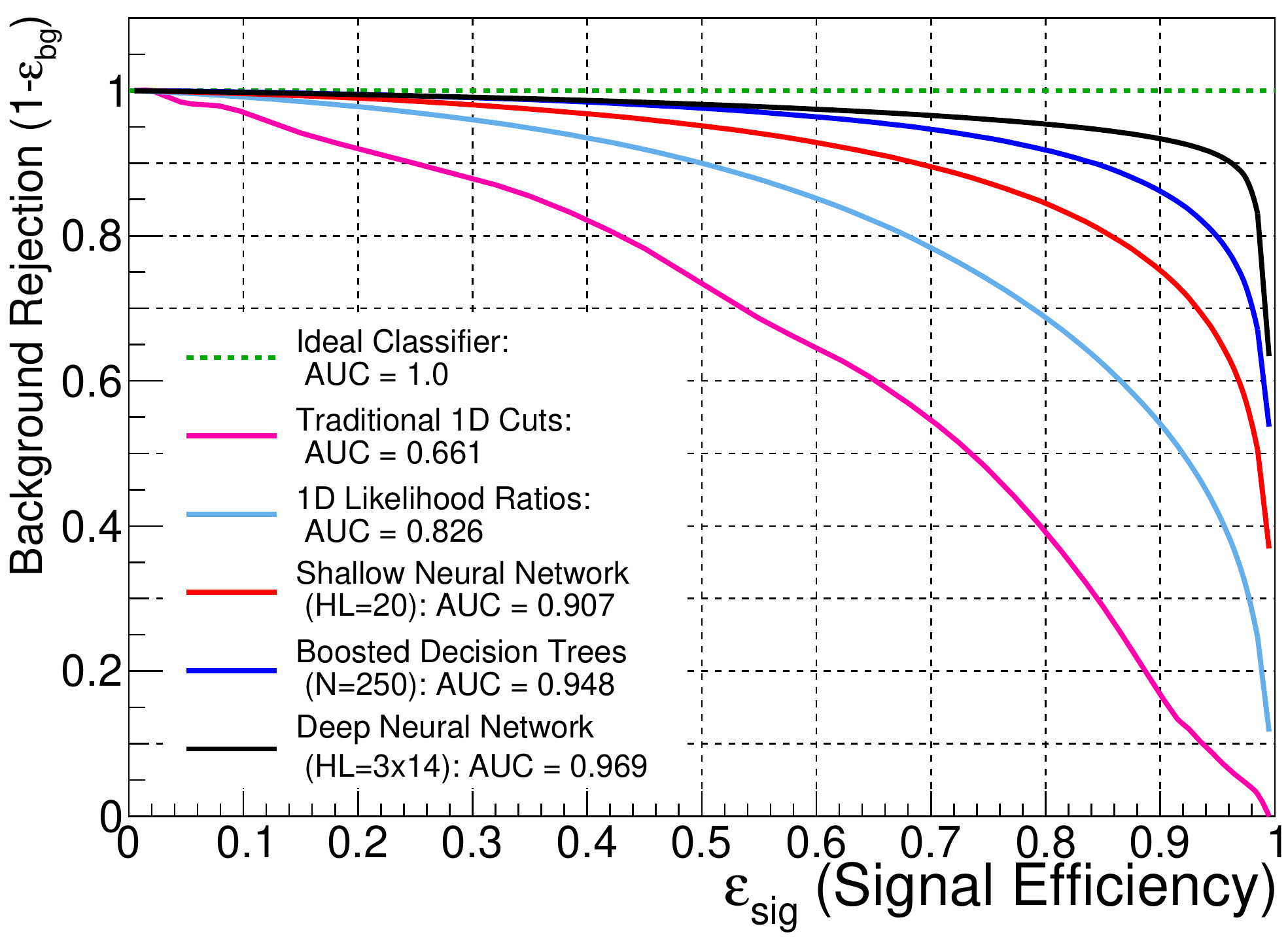}
	\caption{ The background rejection ($1-\varepsilon_{bg}$) versus the signal efficiency ($\varepsilon_{sig}$) for several different multivariate classifiers and traditional 1D cuts. }
	\label{fig:roc}
\end{figure}

\section{Results and Applications} \label{sec:05results}
\subsection{Comparison of Multivariate Classifiers} \label{sec:0501mvacomp}
In the previous section neural networks were trained as classifiers for the purpose of separating signal muons from various background sources. The performance of the neural network based classifiers are compared using modified receiver operating characteristic (ROC) curves in Fig.~\ref{fig:roc} by plotting the background rejection power ($1-\varepsilon_\mathrm{bg}$) vs.\ the signal efficiency ($\varepsilon_\mathrm{sig}$). The performance of a classifier can be succinctly summarized with the area under the curve (AUC) of the background rejection vs.\ signal efficiency curve. An ideal classifier is able to reject 100\% of the background while providing 100\% signal efficiency and has an AUC of 1. On the other hand, a random guess classifier has an should have a 50/50 chance of correctly guessing the class and has an AUC of 0.5. 

The neural network classifiers shown in Fig.~\ref{fig:roc} are also compared with classifiers employing optimized 1D cuts, 1D likelihood ratios, and boosted decision trees (BDTs). The cuts used in the 1D cut classifier were optimized on the $J\psi$ peak in $p$+$p$ collisions at $\sqrt{s}$ = 200 GeV. Both the 1D likelihood ratio classifier and the BDTs were trained using the TMVA package. The 1D likelihood ratio classifier was trained with default parameters using spline interpolation when building the feature PDFs. The track $p_T$ and charge ($q$) variables were removed from the 1D likelihood classifiers since they should not be used directly for muon identification. Additionally, since 1D likelihoods cannot properly incorporate the $p_T$ dependence of the $\Delta$TOF, $\Delta Y$, and $\Delta Z$ features, the 1D likelihood classifier was evaluated only for tracks in a narrow $p_T$ range (1.4~$< p_T <$~1.6 GeV/$c$). A more thorough look at using likelihood ratios for muon identification with the MTD can be found in \cite{Huang2016}. The BDT classifier was trained with $NTrees=250$ and $MaxDepth=5$ with all other parameters set to the defaults. 

\begin{figure}
	\centering
    \includegraphics[width=0.7000\textwidth]{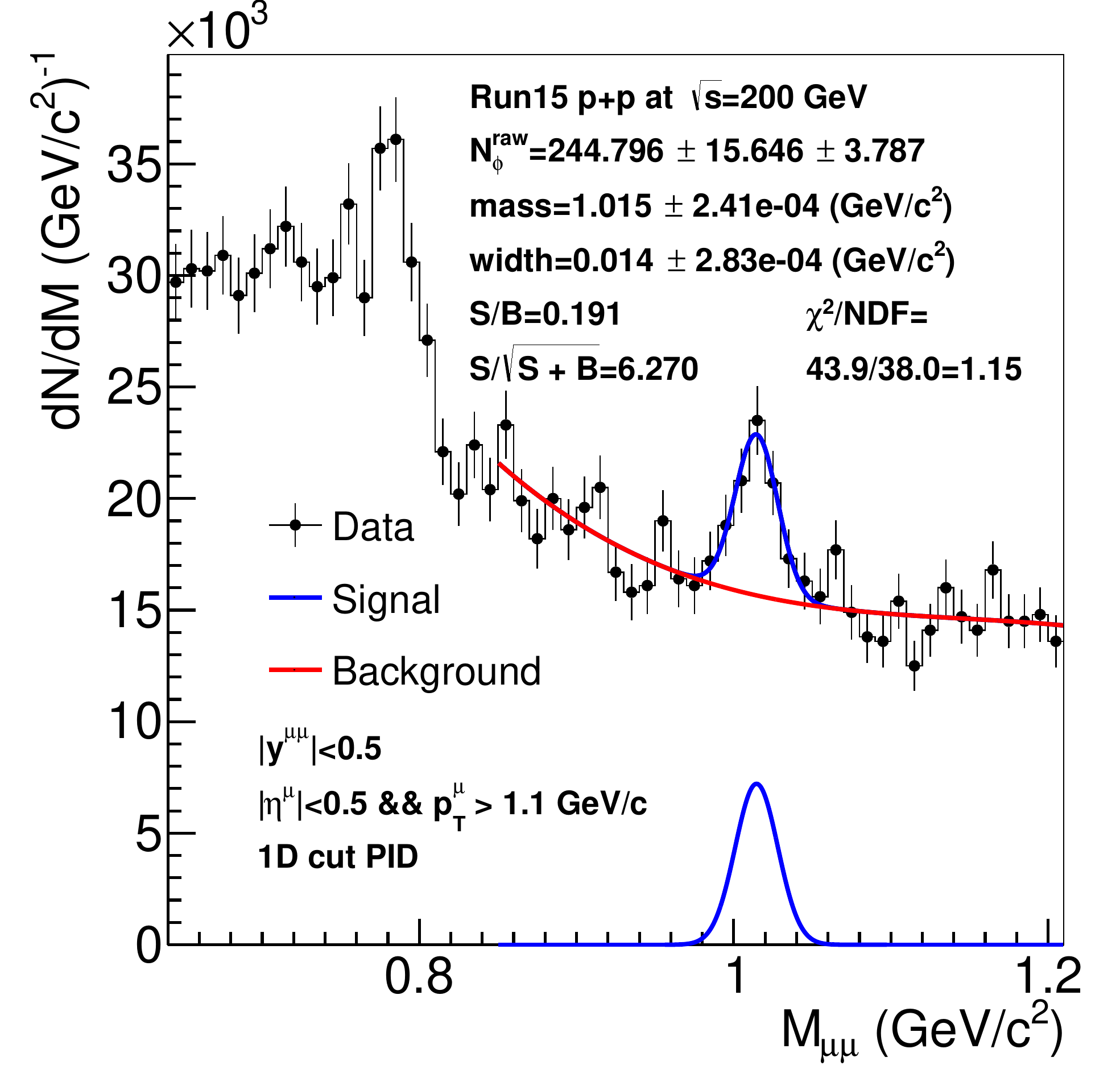}\label{fig:phi-1}
	 \caption{ Raw yield extraction of the $\phi$ meson using optimized traditional 1D PID techniques. }
	\label{fig:phi}
\end{figure}

\begin{figure}
	\centering
    \includegraphics[width=0.7000\textwidth]{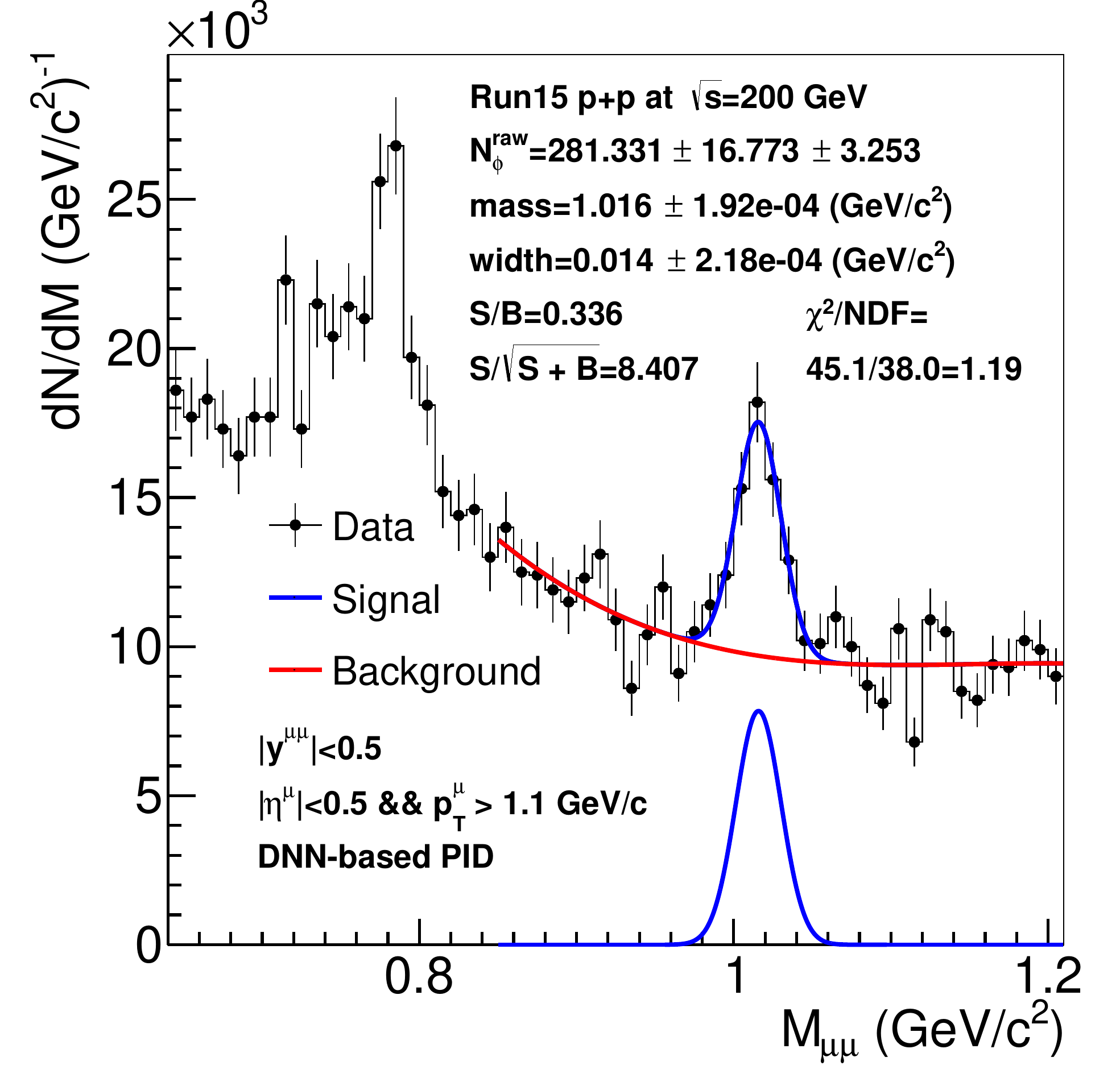}\label{fig:phi-2}
	 \caption{ Raw yield extraction of the $\phi$ meson using the DNN based PID. }
	\label{fig:phi}
\end{figure}

\subsection{Muon Identification in Data}
 The DNN classifier out-performed the other multivariate classifiers investigated in Sec.~\ref{sec:04muonid} based on an analysis of the background rejection power vs.\ signal efficiency evaluated on a testing sample of simulated events. We can further test the performance of the DNN classifier by applying it to the dimuon data collected from $p$+$p$ collisions at $\sqrt{s}=200$~GeV. The decay of resonances to muons, like the $\phi \rightarrow \mu^+\mu^-$ decay, provides a self-analyzing set of data for testing muon identification techniques. Muon pairs are selected in the data by first evaluating the DNN response for all muon candidates in an event. Pairs are then formed from oppositely charged muons. Signal pairs are selected based on the pair DNN response $r_\mathrm{pair}$:
 \begin{equation}
     r_\mathrm{pair} = \sqrt{r_a^2 + r_b^2}
 \end{equation}
 where $r_a$ and $r_b$ are the DNN responses for paired muons $a$ and $b$, respectively. The DNN was specifically optimized to promote a response of $r\approx1$ for signal muons and a response of $r\approx0$ for background sources. Consequently, the pair response for a $\mu^+\mu^-$ pair will be $r_\mathrm{pair}\approx\sqrt{2}$. The optimal $r_\mathrm{pair}$ cut for selecting $\phi\rightarrow \mu^+\mu^-$ decays was determined by maximizing the $\phi$ significance ($S/\sqrt{S+B}$) in steps of $r_\mathrm{pair}=0.01$.  The signal and background contributions were extracted by fitting the raw $\mu^+\mu^-$ invariant mass spectra in 0.85 $ < M_{\mu\mu} < $ 1.5 GeV/$c^2$. A 4$^{\rm th}$-order polynomial was used to model the background and a Gaussian was used for the $\phi$ meson peak.  The optimal cut was found to be $r_\mathrm{pair}>1.36$ which provides a $\phi$ meson significance of $\sim$8.3 and a $S/B$ ratio of 0.33. Figure~\ref{fig:phi} shows the raw $\phi$ meson yield extraction fits using traditional 1D cuts optimized on the $J/\psi$ for muon identification and using the DNN-based muon identification. The DNN-based muon identification simultaneously provides higher $S/B$ ratio, significance, and signal efficiency compared to the optimized 1D muon identification. In Fig.~\ref{fig:pidcompare}, the raw $\mu^+\mu^-$ invariant mass spectra in the range 0 $< M_{\mu\mu} <$~4.5 GeV/c$^2$ is shown for optimized 1D cut-based muon identification and compared with the DNN-based muon identification. In addition to improving $S/B$ and significance of the $\omega$ and $\phi$ mesons, the DNN-based muon identification allows the $\psi(2S)$ to be visible.  

\begin{figure*}
	\centering
    \includegraphics[width=0.9500\textwidth]{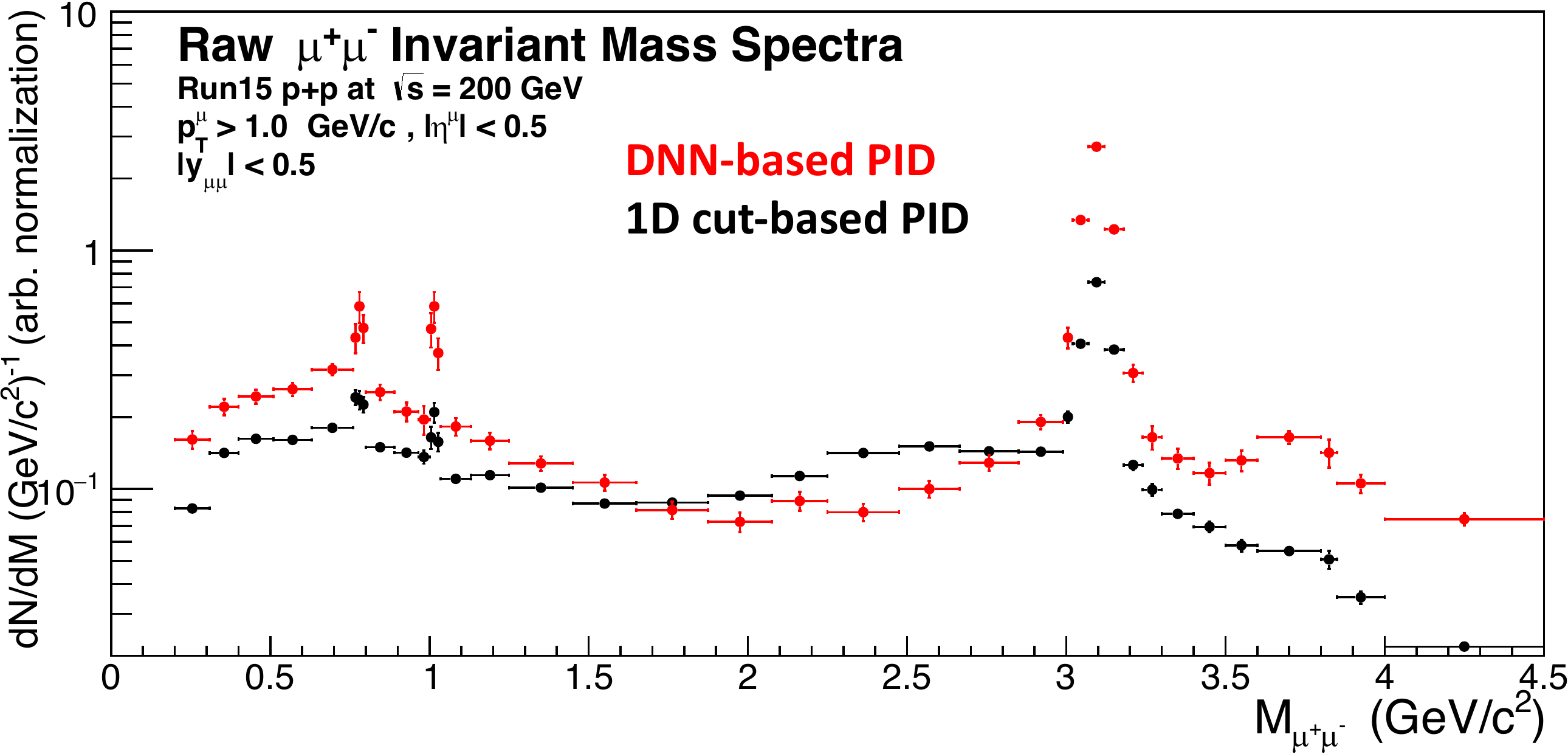}\label{fig:pidcompare}
	 \caption{ Comparison of the raw $M_{\mu\mu}$ invariant mass distribution using optimized 1D cut-based muon identification versus the DNN-based muon identification. The distributions are scaled in 1.5 $< M_{\mu\mu} <$ 2.5~GeV/$c^2$ to make comparison easier. }
	\label{fig:pidcompare}
\end{figure*}

\subsection{Muon Purity Measurements}
Since no individual feature among the set of PID features clearly separates signal from background contributions, it is not possible to fit any one of the features in order to extract the muon purity of tracks in data. Given the signal and background PDFs for each of the 8 PID features (neglecting $p_T$ and $q)$, one could in principle conduct a simultaneous fit to all 8 distributions in order to extract the yield of signal and background contributions. Since each distribution would need to be fit to a $\mu$, $\pi$, $K$, and $p$ contribution it would require simultaneously fitting 8 distributions with 32 templates constrained by 4 free yield parameters. While possible, in practice a simultaneous fit with so many distributions and templates is technically challenging and often proves unstable. 

\begin{figure}[H]
	\centering
    \includegraphics[width=0.60\textwidth]{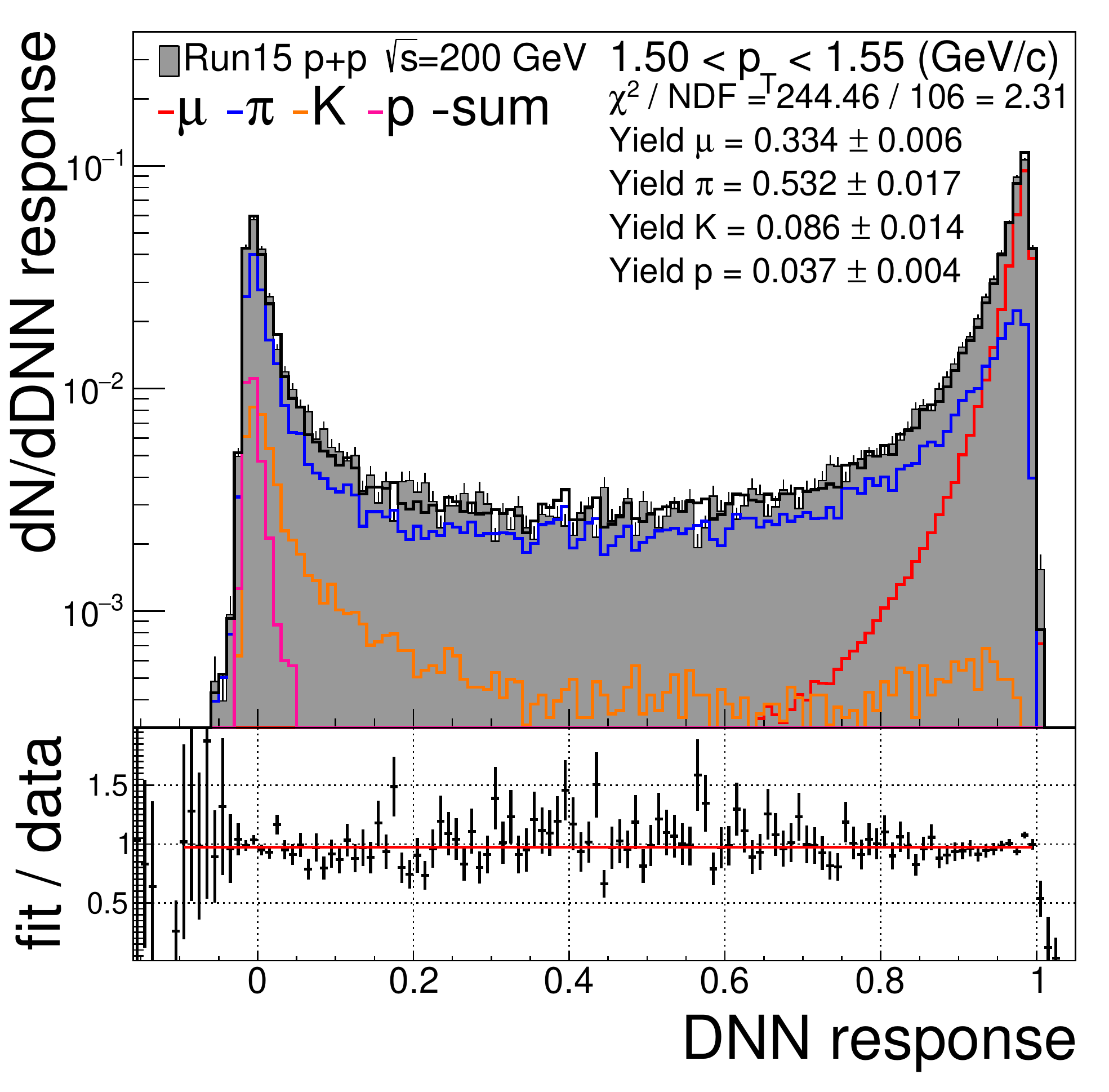}
	 \caption{ The top panel shows the DNN response for muon candidates in the range 1.5 $< p_T <$ 1.6~GeV/$c$. A template fit is conducted to extract the contributions from $\mu$ (red), $\pi$ (blue), $K$ (orange), and $p$ (magenta). The lower panel shows the ratio of the data over the sum of the contributions.}
	\label{fig:purityfit}
\end{figure}

Instead, the complexity of the problem can be greatly reduced by simply fitting the DNN response for muon candidates with the template shapes for signal and background components. Since the DNN combines all PID features  In this setup, only a single distribution needs to be fit with the 4 template shapes for signal and background each with a free yield parameter. Figure~\ref{fig:purityfit} shows the result of this procedure applied to muon candidate tracks in the range 1.5 $< p_T <$ 1.6~GeV/$c$. The template for each component is computed by evaluating the DNN on simulated tracks in the same kinematic regions as those in the data. The data/fit ratio shown in the lower panel of Fig. \ref{fig:purityfit} shows that the fit is capable of describing the DNN response for muon candidates to within $\sim$20\% over the entire range of DNN responses. 

\begin{figure}[H]
	\centering
    \subfloat[]{\includegraphics[width=0.48\textwidth]{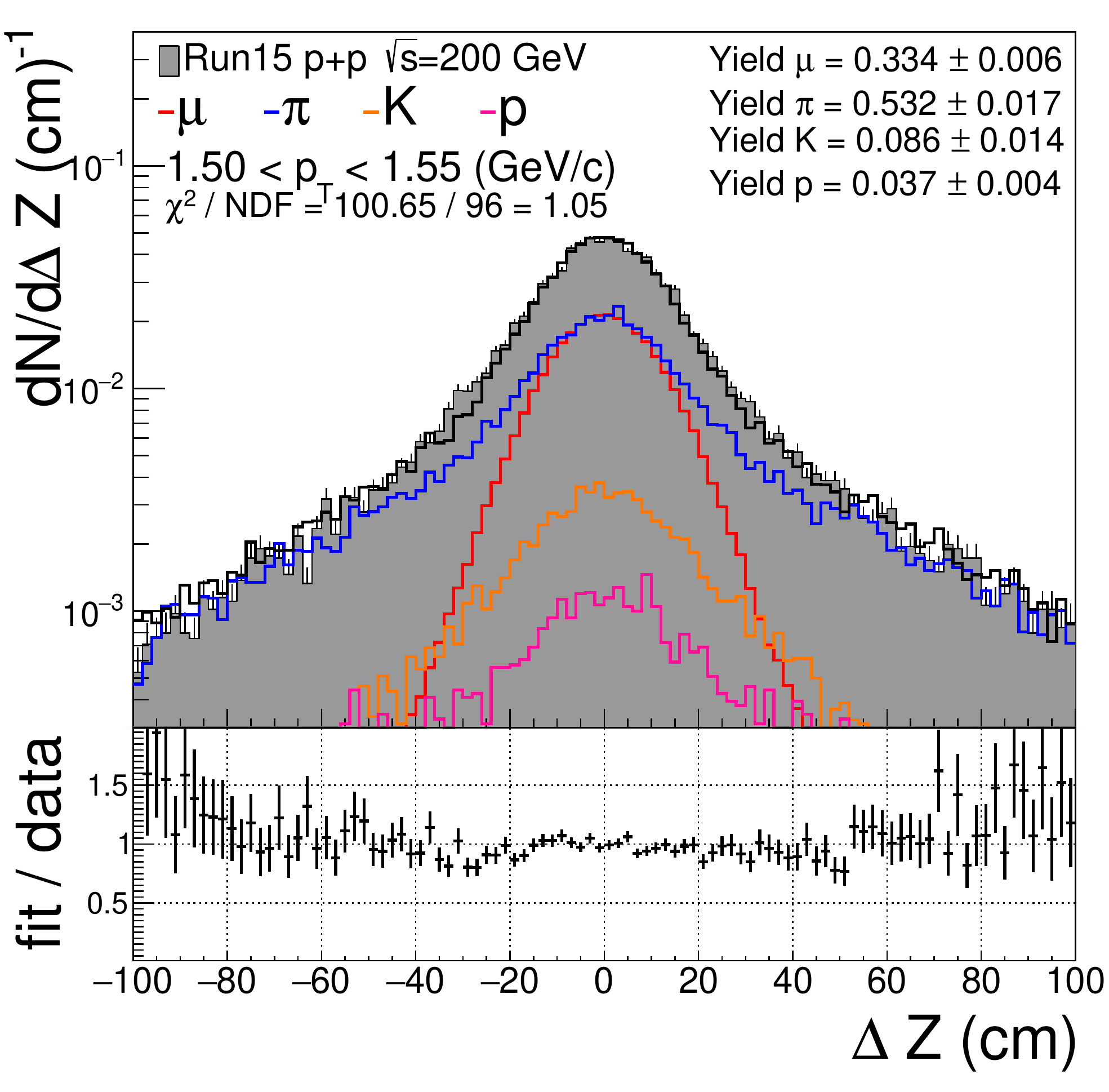}\label{fig:fitdZ}}
    \subfloat[]{\includegraphics[width=0.48\textwidth]{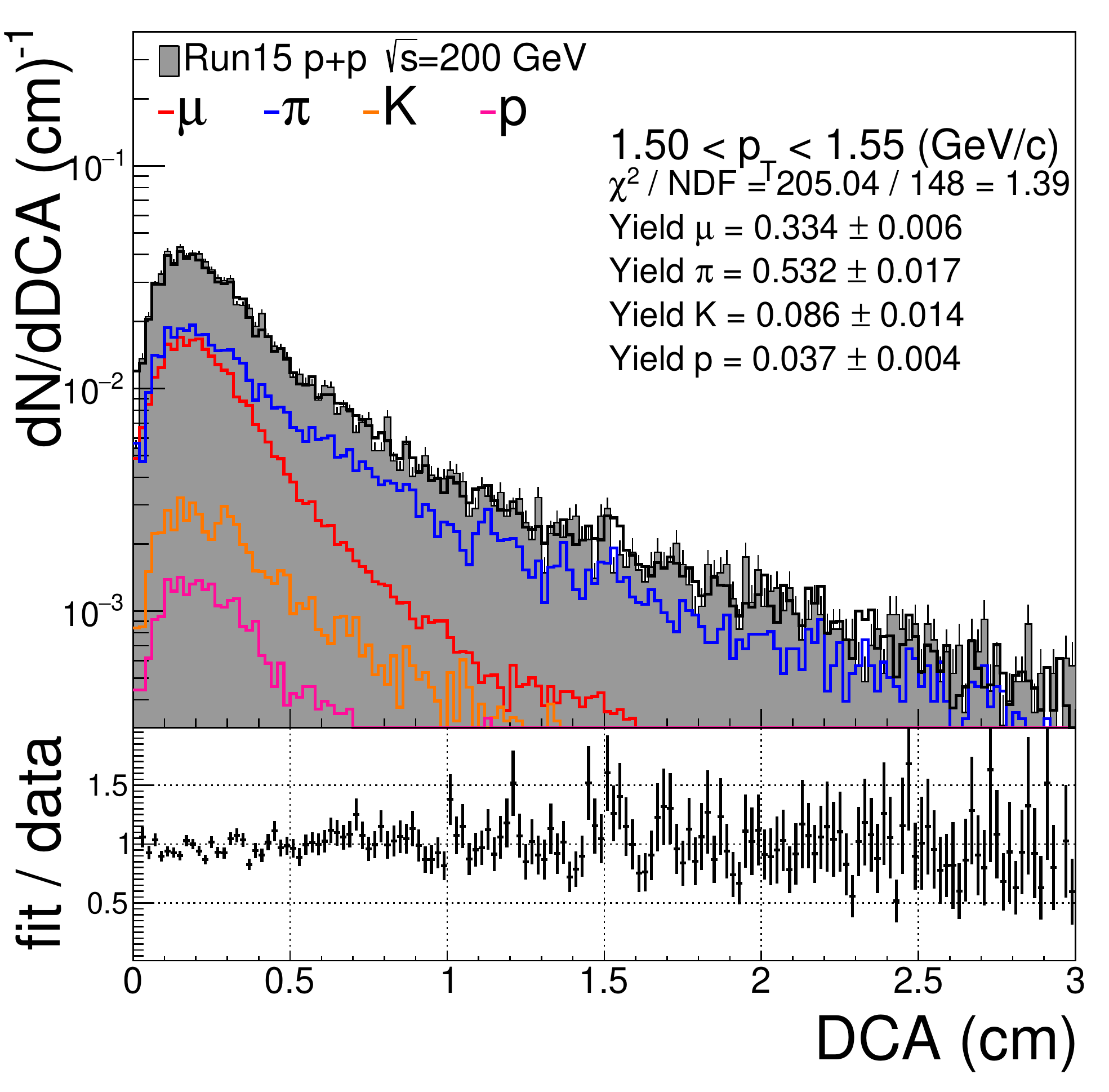}\label{fig:fitdca}}
	 \caption{ The result of the DNN response fit for $\mu$, $\pi$, $K$, and p contributions projected back onto the $\Delta Z$ (a) and DCA (b) distributions. The ratio of fit over data is shown in the lower panels of each figure. }
	\label{fig:fitprojections}
\end{figure}

After determining the yield of each signal and background contribution, the DNN response can be projected back onto all of the 8 PID features to verify that the DNN is properly combining the information from all variables. Ensuring that the projection onto each PID feature results in a good description of the data is a strong demonstration that the DNN is not over-training on artifacts in the training samples. Projections onto the $\Delta Z$ and DCA features are shown in Figs.~\ref{fig:fitdZ} and \ref{fig:fitdca}. This technique allows the increased signal vs. background separation power provided by the DNN-based muon identification to be leveraged for data-driven muon purity measurements. At the same time, the ability to project the muon purity fit results back onto the PID features provides a data-driven strategy to test for over-training and poor model generalization.

\section{ Summary } \label{sec:06summary}  
The installation of the muon telescope detector has made muon identification possible at STAR over a large $p_T$ range. With only a single layer of steel acting as a hadron absorber, backgrounds from hadron punch through and weak decays make primary muon identification challenging. Several quantities measured by the STAR tracker and MTD are used to train shallow and deep neural network classifiers for the purpose of muon identification. The deep neural network classifier out-performed the other multivariate classifiers investigated in Sec. \ref{sec:04muonid} based on an analysis of the background rejection power vs. signal efficiency evaluated on a testing sample of simulated events. When applied to dimuon triggered p+p collisions at $\sqrt{s} = 200$~GeV, the DNN-based PID simultaneously provides higher $S/B$ ratio, significance and efficiency for the $\phi$-meson yield extraction. At higher masses, the he DNN-based muon identification makes the $\psi(2S)$ state significantly more visible in the raw $M_{\mu\mu}$ distribution compared to optimized 1D cut-based muon identification. Finally, an application of the trained DNN for data-driven muon purity measurements is presented. 

\section{Acknowledgements}\label{sec:ack}
We thank the STAR Collaboration for the use of the experimental data
shown in this paper and the operation of this system during RHIC running
periods as part of STAR standard shift crew operations. This work was funded by the U.S.\ DOE Office of Science under contract No.~DE-FG02-10ER41666.



 \bibliographystyle{elsarticle-num} 





\end{document}